# How "Short" a "Short-term earthquake prediction" can be? A review of the case of Skyros Island, Greece, EQ (26/7/2001, Ms = 6.1 R).

Thanassoulas[1], C., Klentos[2], V.

1. Retired from the Institute for Geology and Mineral Exploration (IGME), Geophysical Department, Athens, Greece.
   e-mail: thandin@otenet.gr - URL: www.earthquakeprediction.gr

2. Athens Water Supply & Sewerage Company (EYDAP),
   e-mail: klenvas@mycosmos.gr - URL: www.earthquakeprediction.gr

**Abstract.**

The earthquake of Skyros Island in Greece (26/7/2001, Ms = 6.1 R) was announced three days before its occurrence in a meeting organized by the Bulgarian Academy of Sciences, Sofia on 2001. In this work a review is made on the results of the methodologies which were used for the analysis of the available data in order to utilize the specific prognosis. Of particular interest is the fact that the determined time of occurrence deviated for less than an hour (actually it was 41 minutes) from the actual occurrence time of the specific EQ. A similar example of another large EQ (Kythira, Greece, Ms = 6.9R, 8th January, 2006) by using the very same methodology deviated for 43 minutes only. The rest of the prognostic parameters: the epicenter location and the magnitude of the Skyros EQ were utilized by manipulating the past seismic energy release history of the regional seismogenic area and the observed earth's electric field which was recorded during the last 25 days, before the occurrence of the large seismic event, by VOL earth's electric field monitoring site. It is concluded that a short-term earthquake prediction can be utilized as a really "short" one, once the available precursory data are processed with the appropriate methodologies.

## 1. Introduction.

The term "earthquake prediction" means that the prognostic parameters of an earthquake (a large one) which are: the epicenter, the time of occurrence and the magnitude will be determined by any suitable methodology, "a priori". Moreover, according to the estimated prognostic time window of its occurrence, the "earthquake prediction" is characterized as: long term, when the time window is of the order of some decades of years, medium term when the time window is of the order of some years or short-term for shorter time windows.

By taking into account the fact that usually large earthquakes cause large damages in technical constructions and quite often human deaths (recent Haiti earthquake) it is evident that it is of great importance the implementation of a short-term earthquake prediction scheme. Furthermore, the specific prediction must be put in effect at some time before the seismic event, so that there is enough time to take some practical actions against it, while it is very important not to mobilize the society for a long period before its occurrence and thus to create more social panic and chaos than an unexpected seismic event could create.

Consequently, the raised question is: how "short" a short-term earthquake prediction can be? This question will be answered for the case of Skyros Island, Greece, EQ (July 26th 2001, Ms = 6.1R). This particular earthquake was announced, three days before its occurrence, in a seminar on earthquake prediction organized by the Bulgarian Academy of Sciences, held in Sofia, Bulgaria on 23rd – 26th July, 2001 (Thanassoulas, 2001).

The methods which were used for this prediction have been presented in details in a monograph under the title "Short-term Earthquake Prediction" (Thanassoulas, 2007). Those who are interested in the details of the methodologies can download, in PDF form, the entire monograph from the site: www.earthquakeprediction.gr

It must be remembered too, that large earthquakes are not so frequent, thus prohibiting a real statistical test of the obtained deterministic results. Moreover, the generating mechanism, in each large earthquake, is more or less different from any other that takes place at a different seismogenic area. Consequently, whatever is applicable to one large earthquake may not be applicable to another one. Even if this is the case, it is worth to predict just only one earthquake and to save human lives than to ignore it and accept human losses as an unavoidable fact.

## 2. Data presentation and analysis.

The data (earth's electric field registrations) which are used in this work were acquired by VOL monitoring site, run and maintained by John Tsatsaragos, during our collaboration from 1999 to 2002. The data were acquired by a set of two dipoles of about 110m length each and oriented approximately to N-S and E-W directions. The data were normalized to equal length dipoles and to the true E-W and N-S directions. The location of the VOL monitoring site in relation to the Skyros EQ epicenter is shown in the following figure (1).

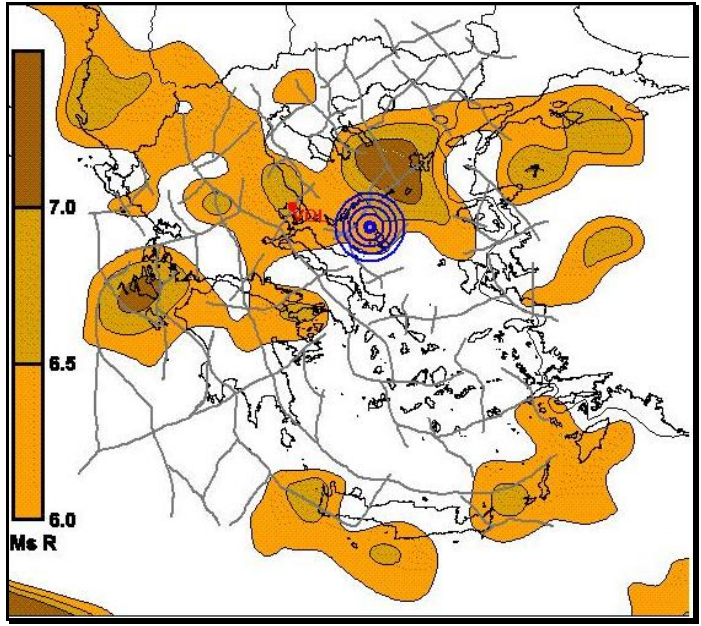

**Fig. 1.** Location of VOL monitoring site (red lettering) and Skyros EQ epicenter (concentric blue circles). Thick gray lines represent the lithospheric deep fracture zones / faults while the brown colored areas indicate the stress charged areas capable to generate an EQ larger than Ms = 6.0R

The background of figure (1) is the seismic potential of the Greek territory for the year 2000 (Thanassoulas et al. 2003)

**Skyros raw data.**

The normalized earth's electric field data are presented in the following figure (2). The data spans from the 1st of July to the 8th of August 2001. The red bar indicates the time of occurrence of the Skyros EQ.

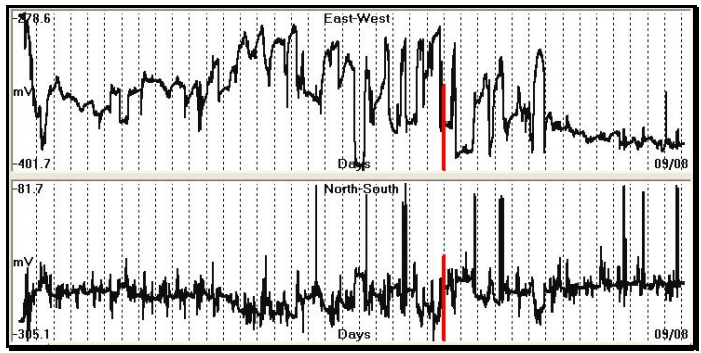

**Fig. 2.** Data of the earth's electric field normalized to E-W and N-S directions, registered by VOL monitoring site. The red bar indicates the time of occurrence of the Skyros EQ.

It is evident from figure (2) that a few days before the main seismic event the earth's electric field was influenced by sharp and large amplitude electric pulses. These pulses last longer in the E-W component while are sharper in the N-S.

**Time determination.**

The first prognostic parameter which will be determined is the time of occurrence. Concurrently with the recording of the raw data, the used active band-pass filter indicates an increasing in amplitude incoming signal of T = 24 hours. This is presented in the following figure (3).

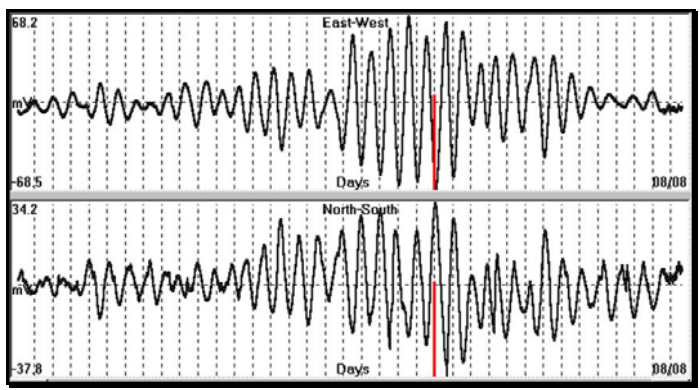

**Fig. 3.** Oscillating component of T = 24hours of the earth's electric field registered by VOL monitoring site observed prior to Skyros, (red bar) earthquake (Ms=6.1R) in Greece (2001).

When the oscillating earth's electric field increases its amplitude it suggests that a seismogenic area has reached the critical stress charge level which happens a short time before the occurrence of the future seismic event (Thanassoulas 1982, 2001, 2007, Thanassoulas et al. 1986, 1993, 2008). The seismogenic area under such stress load conditions generates (through the activation of a large scale piezoelectric mechanism, Thanassoulas 2008) an anomalous preseismic electric field (fig. 2) of which the monochromatic of T = 24 hours component is identified by the active band-pass filter.

At this stage we know that an earthquake is pending but we have no clue yet about its occurrence time. What is known is the fact that a slight increase of the stress in the seismogenic area can trigger the pending seismic event. The time of triggering can be determined (Thanassoulas, 2007) by the well known tidal waves and the corresponding oscillating stress component combined (Boolean AND) with the time of the generation of the earth's electric anomalous preseismic signals. Consequently, we apply the following scheme shown in figure (4).

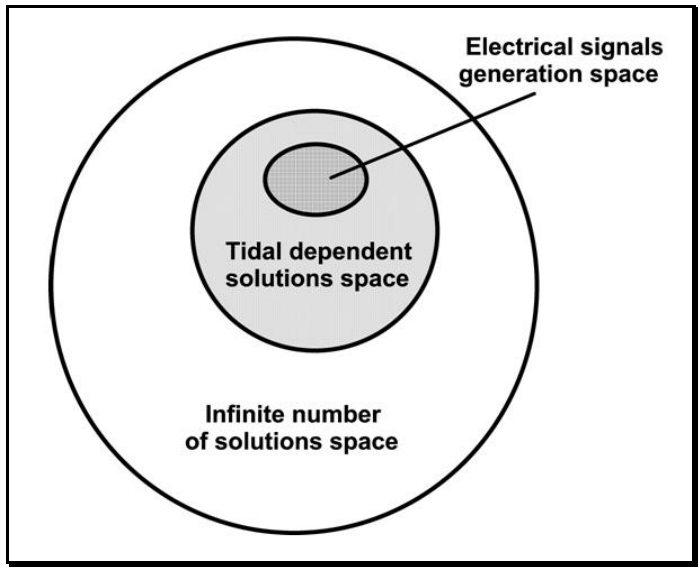

Fig. 4. Time of occurrence of a large EQ determination schematic presentation by using Boolean algebra (after Thanassoulas, 2007).

At first we consider the T = 14 days period tidal wave presented in the following figure (5). The red bar indicates the time of occurrence of Skyros EQ.

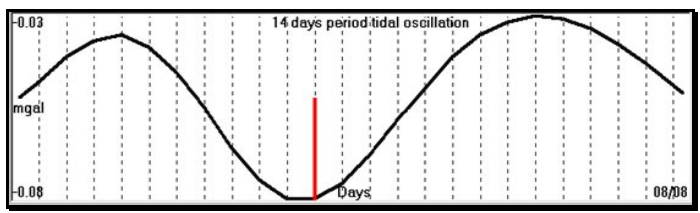

Fig. 5. Lithospheric oscillation of T = 14 days (and corresponding stress load) vs. time of occurrence of the EQ (red bar) of Skyros, Greece, 26/7/2001, Ms = 6.1 R.

It has been shown (Thanassoulas 2007) that the majority of the large EQs do occur on the peaks of the tidal oscillations. Figure (5) represents exactly this case. The Skyros EQ did occur on the peak of the 14 days period tidal wave, but yet we do not know it. The next step is to combine the corresponding tidal wave with the oscillating component of the earth's electric field (Boolean AND operation). This is presented in the following figure (6).

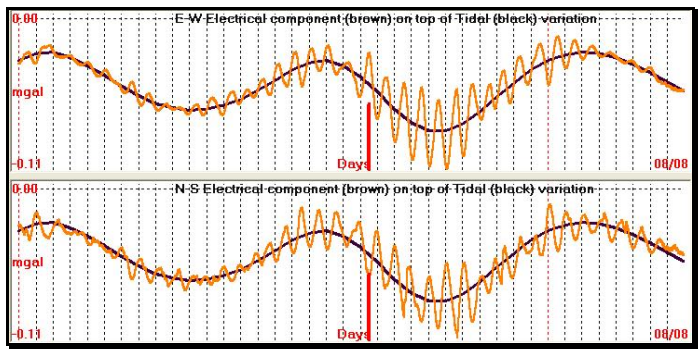

Fig. 6. Earth's oscillating electric field (T = 24 hours, brown line) on top of the tidal wave (T = 14 days, black line). The red bar indicates a fore shock of Ms = 5.1R which occurred on 21st of July 2001.

Two characteristic features are important in figure (6). The first is that an earthquake of Ms = 5.1R takes place on the 21st of July, 2001 and the earth's oscillating electric field does not decrease in amplitude as it would be expected to if the seismogenic area had been stress load discharged. It is evident that the seismogenic area is still under critical stress load conditions and capable to generate a larger EQ since the amplitude of the earth's oscillating field steadily continues to increase.

The second characteristic is the time of occurrence of the next to come maximum stress load (25 – 26 of July, 2001). This is well known, in advance, since the tidal wave peaks (T = 14 days) take place at particular times for any time period, and for any location of specific geographical coordinates, as calculated by a FORTRAN software (Rudman et al. 1977).

Consequently, the 25$^{th}$ and 26$^{th}$ of July are the most probable days, when the pending EQ will occur, since during these days, due to tidal wave peak presence, the stress load acquires a maximum value. Moreover, the location of the foreshock suggests the regional seismogenic area that has been activated. In the following figure (7) the location of the foreshock is shown in relation to the VOL monitoring site location.

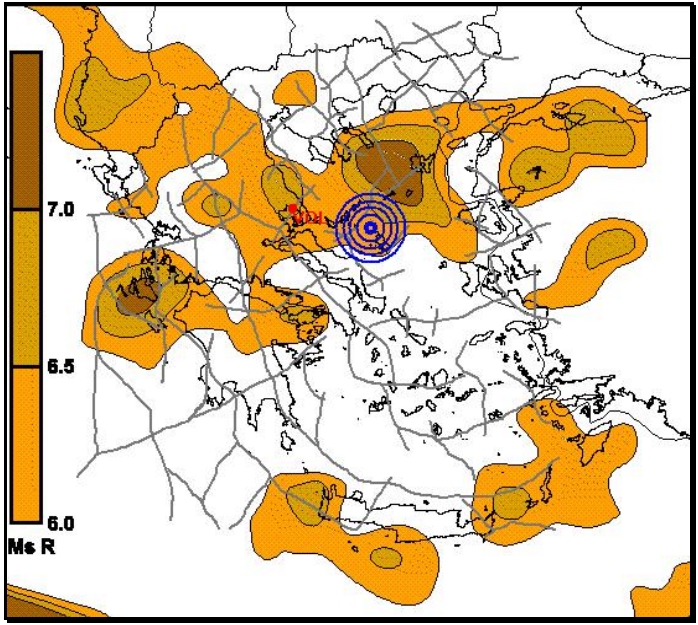

**Fig. 7. Foreshock location (blue concentric circles) is presented in relation to the VOL (red lettering) location.**

The main event (Ms = 6.1R) actually did occur on 26$^{th}$ of July, 2001, 00h 21m GMT. The following figure (8) prescribes the oscillating earth's electric field in relation to the tidal wave (T=14 days) and the time of occurrence of the main seismic event.

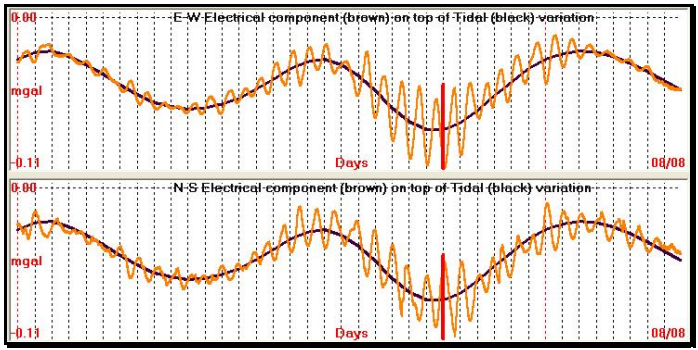

**Fig. 8. Earth's oscillating electric field (brown line) in relation to the tidal wave (black line) and the time of occurrence (red bar) of the main seismic event.**

The first thing to notice in figure (8) is the fact that the pending EQ did occur, as it was expected, at the peak of the tidal wave. The second is that the earth's electric field oscillation amplitude decreased drastically after the seismic event. Therefore it is concluded that the seismogenic area had been, at a certain level, stress discharged and this seismic event was the main one. Although at this stage we new the oncoming occurrence of the pending EQ for some days before, the time of its occurrence was estimated within a time window of a couple of days. Therefore, in order to determine a shorter time window for its occurrence, the specific period of 25$^{th}$ – 26$^{th}$ of July, which corresponds to the tidal peak of the 14 days period tidal wave, was studied in terms of the daily tidal variation. The idea behind this analysis is that the seismic event, finally, will be triggered by a very small stress increase as long as it is just before final triggering. This last critical final stress increase is provided within a very short period of time by the daily tidal oscillation. The following figure (9) shows the case for Skyros EQ.

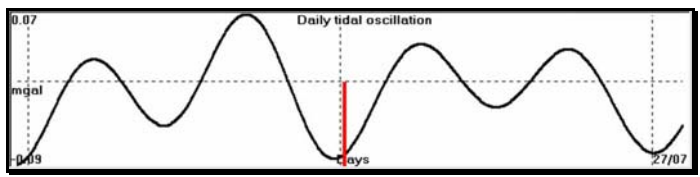

**Fig. 9. Daily, tidal, lithospheric oscillation (black line) vs. time of occurrence (red bar) of the EQ in Skyros, Greece, 26/7/2001, Ms = 6.1 R.**

The suggested time of the EQ occurrence from figure (9) is: 23hours and 40minutes GMT of the 25th of July, 2001. The Skyros EQ did actually occur on 00hours and 21minutes GMT of the 26th of July, 2001. The difference between the suggested time of occurrence and the actual one is only 41 minutes.

A very similar case is the one of Kythira, Greece EQ (Ms = 6.9R, 8th January, 2006). In the following figures (10) and (11) are presented the time of occurrence of the large EQ in relation to the tidal wave of T=14 days (fig. 10) and in relation to the daily tidal variation (fig. 11) for the 7th, 8th and 9th of January of 2001.

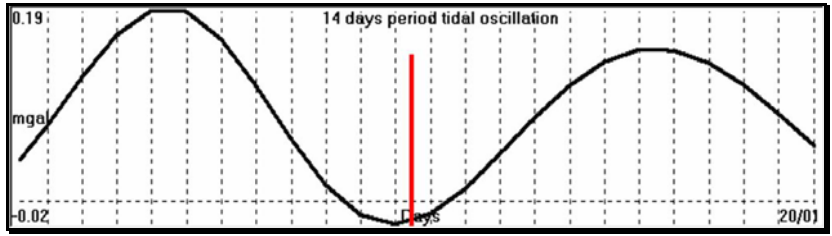

Fig. 10. Time of occurrence of the Kythira EQ (red bar, Ms = 6.9R, 8th January, 2006) in relation to the tidal wave (black line) of T = 14 days.

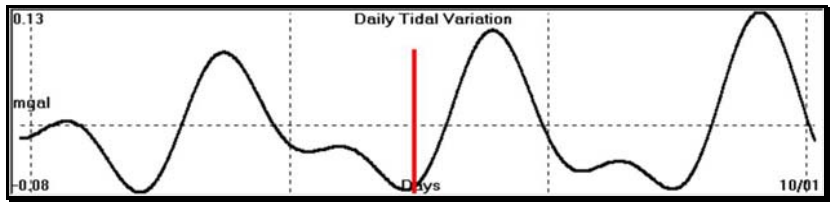

Fig. 11. Time of occurrence of the Kythira EQ (red bar, Ms = 6.9R, 8th January, 2006) in relation to the daily tidal variation (black line) of T = 1 day.

It is concluded from figure (10) that the most probable days for the occurrence of the pending EQ are the dates of 7th and 8th of January of 2006. Furthermore, figure (11) which is the detailed daily variation of the tidal wave for the dates 7th, 8th and 9th of January suggests a time of occurrence as follows:

a) for the 7th of January : time of occurrence = 10hours 09minutes GMT

b) for the 8th of January : time of occurrence = 10hours 51minutes GMT

The Kythira EQ did occur on the 8th of January of 2006 on 11hours 34minutes GMT. Therefore, the difference between the actual time of its occurrence and the suggested one for the 8th of January is only 43 minutes.

**Epicenter determination.**

The second prognostic parameter to be determined is the location of the epicenter of the expected seismic event. In the case of Skyros EQ only one monitoring site is used and therefore, only the azimuthal direction of the epicenter location in respect to the monitoring site location can be estimated. A hint concerning the activated regional seismogenic area was given by the occurrence of the foreshock of Ms = 5.1R of the 21st July, 2001. In the following figures (12 – 17) the determination of the azimuthal direction of the foreshock / main shock – VOL monitoring site will be presented by the analysis of the oscillating earth's electric field (Thanassoulas, 1991, 1991a, 2007) before the occurrence of the foreshock – main shock.

**Foreshock of Ms = 5.1R (21st July, 2001).**

In the following figure (12) is presented the earth's oscillating electric field that preceded the foreshock of 21st July with a magnitude of Ms = 5.1R.

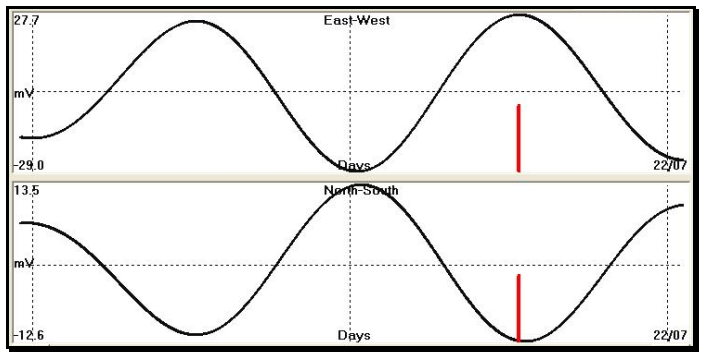

Fig. 12. Earth's oscillating field for the period 20th – 21st July. The foreshock is indicated by a red bar (Ms = 5.1R).

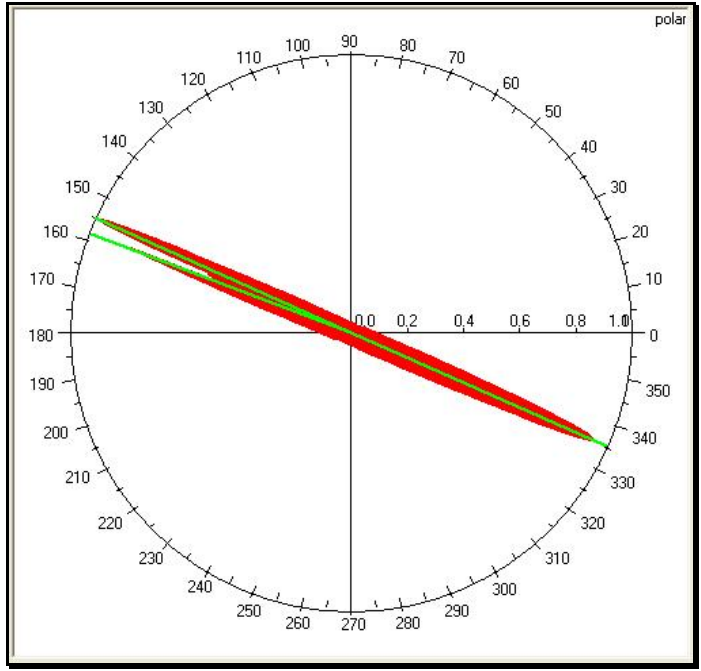

**Fig. 13.** Polar diagram (Thanassoulas et al. 2008) of the earth's electric field intensity vector for the period 20$^{th}$ – 21$^{st}$ July regarding the foreshock (Ms = 5.1R).

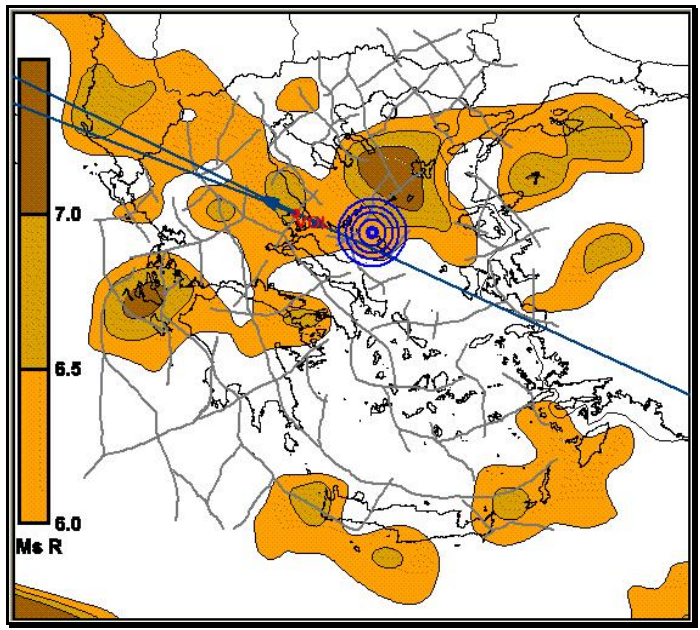

**Fig. 14.** Comparison of the determined azimuthal directions of the earth's electric field intensity vector (blue lines) to the foreshock location (blue concentric circles).

**Main shock of Ms = 6.1R (26$^{th}$ July, 2001).**

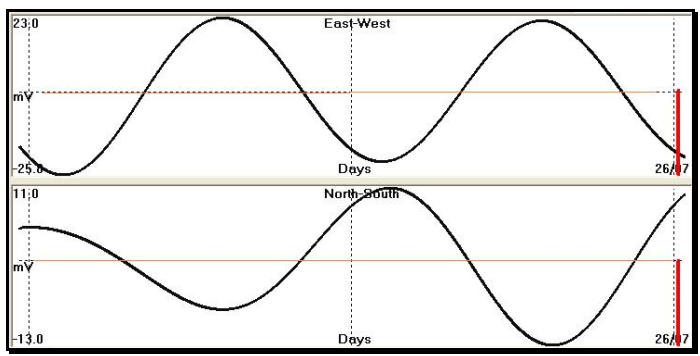

**Fig. 15.** Earth's oscillating field for the period 24$^{th}$ – 25$^{st}$ July. The main shock is indicated by a red bar (Ms = 6.1R).

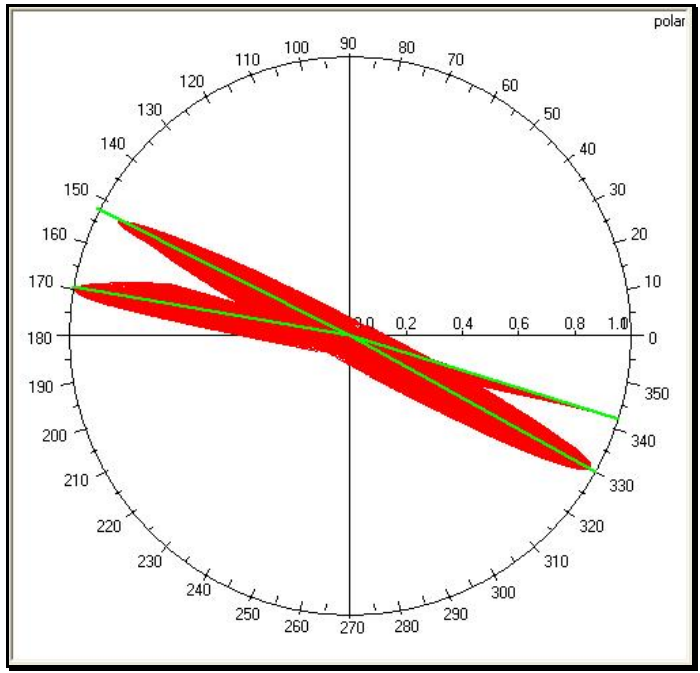

**Fig. 16.** Polar diagram (Thanassoulas et al. 2008) of the earth's electric field intensity vector for the period 24$^{th}$ – 25$^{th}$ July regarding the main shock (Ms = 6.1R).

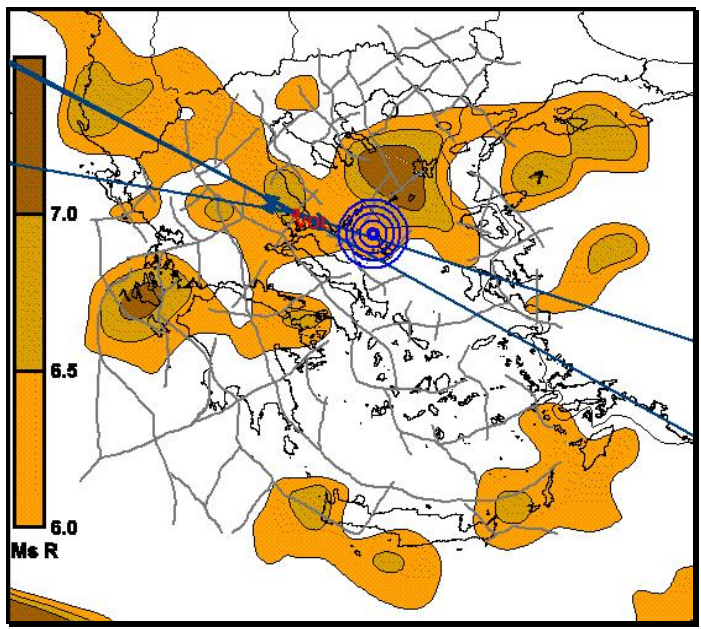

**Fig. 17.** Comparison of the determined azimuthal directions of the earth's electric field intensity vector (blue lines) to the main shock location (blue concentric circles).

**Total earth's electric field.**

Similar determinations can be made for the total earth's electric field (Thanassoulas, 2008) which can be calculated from the original raw data. The corresponding results are presented in the following figures (18 – 20).

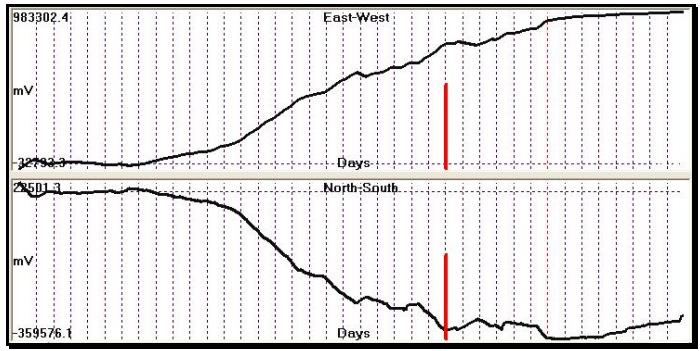

**Fig. 18.** Total earth's electric field calculated from the raw data. The main seismic event (Ms = 6.1 R) is indicated by the red bar.

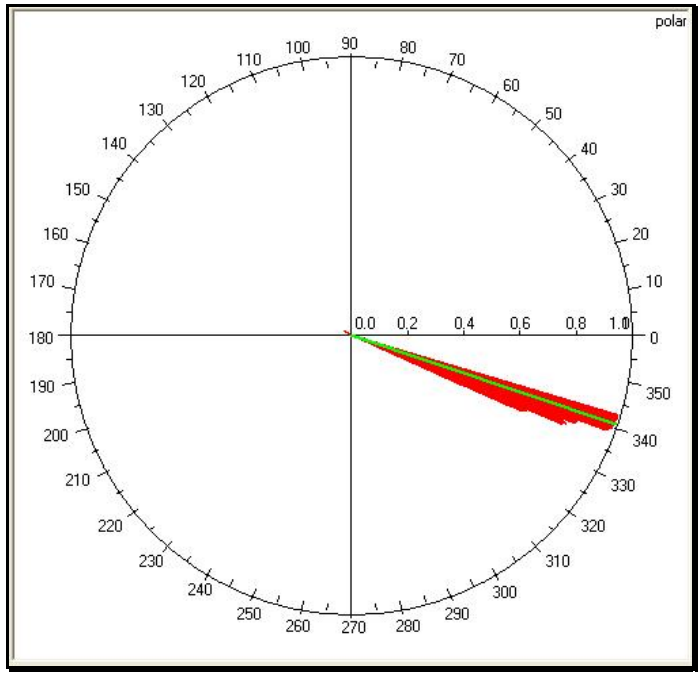

**Fig. 19.** Polar diagram (Thanassoulas et al. 2008) of the earth's electric field intensity vector for the period 1st July – 8th August, regarding the main shock (Ms = 6.1R).

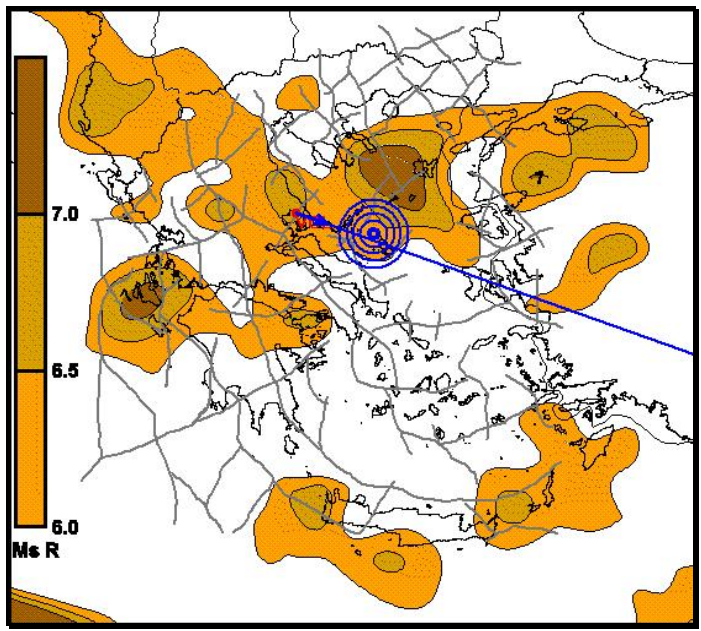

**Fig. 20.** Comparison of the determined azimuthal directions of the earth's electric field intensity vector (blue line) to the main shock location (blue concentric circles).

**Magnitude determination.**

The third prognostic parameter of a pending EQ is its magnitude. For the case of Skyros main EQ after having, in an indirect way, estimated the activated regional seismogenic area, it is possible to apply the "Lithospheric Seismic

**Energy Flow Model" and to determine the maximum expected magnitude (Thanassoulas, 2007, 2008a,b, Thanassoulas et al. 2001.) of the pending EQ. This is demonstrated in the following figure (21).**

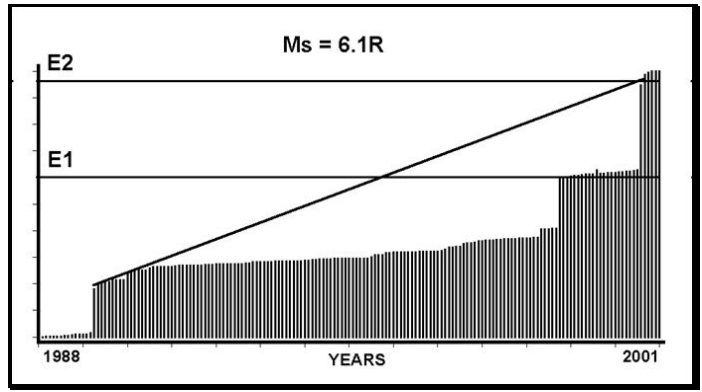

**Fig. 21. Maximum expected magnitude determination (Ms = 6.1R) for the pending Skyros main seismic event.**

### 3. Discussion – Conclusions

In an earthquake prediction procedure, three earthquake parameters must be determined, namely the time of occurrence, the epicenter and the magnitude of the pending earthquake. The accuracy of the parameters determination depends upon the sampling interval of the used data for the corresponding determinations. Although this is quite easy to achieve for the epicenter location and the magnitude, since the data sampling interval is controllable (samples of seismic energy release at any time length, earth's electric field registration at any sampling rate), the situation changes when we try to determine the time of occurrence. The statistical treatment of any large seismic events sequence, at its best, will resolve the time of occurrence with accuracy at a level of sampling interval. If we take into account the fact that large earthquakes are not quite often, then the resolving power of the statistical methods applied on large earthquakes have evidently a limit which is rather small. In other words statistical methods can be applicable to rather medium to long term predictions.

In terms of short-term prediction and because of the inherent drawbacks of the statistical methods, we should follow a quite different approach. From rock mechanics it is known that a rock formation collapses when the applied stress exceeds its fracturing level. In the case of an earthquake the applied stress, short before its fracturing level, is strongly influenced by the excess stress applied through the oscillating lithosphere, triggered by the tidal waves (Thanassoulas, 2007). Therefore, it is of great interest to know in advance the peaks of the oscillating stress at any seismogenic area at any time. This is achieved by the Rudman et al. (1977) FORTRAN program. Consequently, the problem posed now is as follows: at what stress peak a large earthquake will occur? Obviously there is a rather large but limited number of such stress peaks within a year. Moreover, this number varies according to the used tidal wavelength. At this point we have to remind that some time before the occurrence of an earthquake the focal area undergoes drastic changes at its physical properties while a number of mechanisms which generate a preseismic electric field are activated (Thanassoulas, 2007). Thus, by combining the known stress peaks and the generated preseismic electric signals it is possible to narrow the time window within which the pending large earthquake will occur.

The latter is demonstrated in figure (4). The entire procedure has been applied on the Skyros EQ (26/7/2001, Ms = 6.1 R). An anomalous electric field (fig. 2) was detected in July 2001 by the VOL monitoring site. At the same time the active band -pass filter detected an incoming oscillating (T=24h) signal (fig. 3) of increasing amplitude. The latter is an indication that a seismogenic area has reached a critical stress-load level which in turn activates a large scale piezoelectric mechanism. So the next step is to compare the timing of the seismic precursory electric signals with different wavelength tidal components.

The tidal component of T = 14 days is presented in figure (5) in relation to the time of occurrence of the Skyros EQ. It is shown, "a posteriori" at this time that it occurred on the peak of the corresponding tidal wave as the theory predicts (Thanassoulas, 2007).

The earth's electric field oscillating component, in figure (6), is compared to the T=14 days tidal component and to the time of occurrence of a foreshock of Ms = 5.1R which occurred on 21$^{st}$ of July.

What is interesting in figure (6) is the fact that the oscillating electric field didn't decrease in amplitude as it should do if it was the main seismic event. On the contrary, its amplitude continued to increase as it approached the next tidal peak of 25$^{th}$ – 26$^{th}$ of July 2001. Therefore, a more detailed analysis was applied on the tidal wave by using this time a T = 24 hours. The daily tidal variations for the time period of 25$^{th}$ and 26$^{th}$ of July, 2001 are shown in figure (9). The local stress peak of these days is located at 23hours and 40minutes GMT of the 25$^{th}$ of July, 2001 which is the suggested time of occurrence of the pending EQ. The Skyros EQ did actually occur on 00hours and 21minutes GMT of the 26$^{th}$ of July, 2001. The difference between the suggested time of occurrence and the actual one is only 41 minutes.

The case of Skyros earthquake, after presenting this analysis, was announced three days before the occurrence of this seismic event, on the 23$^{rd}$ of July in the INRNE Bulgarian Academy of Sciences during a 3 days seminar that started on 23$^{rd}$ of July, 2001. As an evidence of this presentation, the seminar conclusions were produced as follows.

A very similar case is the one of Kythira, Greece EQ (Ms = 6.9R, 8$^{th}$ January, 2006). The Kythira EQ (fig. 10, 11) did occur on the 8$^{th}$ of January of 2006 on 11hours 34minutes GMT. The difference between the actual time of its occurrence and the suggested one for the 8$^{th}$ of January is only 43 minutes.

## Seminar Conclusions

On the 23rd of July, 2001 a 3-day seminar was held in INRNE, Bulgarian Academy of Sciences, Sofia, Bulgaria, titled:

**"The possible correlation between electromagnetic earth surface fields and future earthquakes."**

During this seminar, **Assoc. Prof. Ranguelov, B., seismologist, Assoc. Prof. Mavrodiev, S. Theor. Phys. and Dr. Thanassoulas, C., Geophysicist** presented their research results on the seminar topic.

Furthermore the aim of the seminar was to investigate the possibilities for submission of a common research program, to establish a regional network for monitoring different geophysical parameters.

The presented examples of measurements of the earth's electric and geomagnetic fields indicate that it is possible to organize such type of a regional Balkan monitoring network for physical and geophysical fields that is going to permit the improvement of the earthquake prediction.

This was validated in practice by the occurrence, during this seminar, in Greece of a large earthquake (Ms=6.1R, 25/07/01) as it had been stated during the presentation of Dr. Thanassoulas and following the corresponding theoretical part of it (time determination), that could happen with very large probability. The magnetic observations, a few days before this large earthquake, presented unusual behavior too, with the specific daily variation of the 25th of July.

To this end the scientific community in Balkan region will be asked to collaborate into the implementation and submission of this project for funding by any appropriate authorities.

For easing the utilization of all the afore mentioned ideas the Organizing Committee proposes the foundation of a non government organization (NGO) under the title:

**"The Balkans, Black Sea Region Sustainable Development (Harmonic Existence) and Science."**

For the INRNE  
Bulgarian Academy of Science

Prof. D.Sc. Stamenov, J.  
/Director/

The Organizing Committee

Assoc. Prof. Mavrodiev, S.Cht. (Bulgaria)

Dr. Thanassoulas, Geophysicist (Greece)

27 July, 2001  
Sofia, Bulgaria

Therefore, we believe that the presence of two different cases of large EQs which behaved in the same way regarding their triggering mechanism (tidal waves) overcomes the case of a by chance incidents, while a more general study of this triggering mechanism (Thanassoulas, 2007) has shown that the vast majority of the large EQs are triggered by the very same lithospheric oscillation tidal mechanism.

The parameter of the epicenter location was dealt with the methodologies presented by Thanassoulas (1991, 1991a, 2007). Although only one monitoring site is available and triangulation is impossible, the regional seismically activated area was indirectly indicated by the presence of the foreshock of 21st of July 2001.

In figures (12 – 17) are presented the earth's oscillating electric field, the corresponding polar diagram and the azimuthal direction of the earth's electric field intensity vector in relation to the location of both the foreshock and the main shock of Skyros EQ.

The same methodology but this time on the total earth's electric field (Thanassoulas, 2008) was applied and is presented in figures (18, 19 and 20). In this case it is shown that the earth's total electric field provides a more stable solution regarding the determined azimuthal direction in relation to the epicenter location of the main seismic event. Moreover, this solution is stable for the entire study period (39 days). The latter is explained by the fact that the total field is produced by integrating the raw data in time and therefore this operation acts as a low-pass filter applied upon the data. On the contrary, since the recorded raw data are in practice an earth's potential gradient (in time and distance from the epicenter) these data are affected by interfering high frequency noise since the gradient operation is equivalent to a high-pass filter.

Another interesting feature of the observed preseismic earth's electric field is its relation to its tidally triggered oscillating component. In figure (22) is presented in a sketch drawing a typical stress / strain curve and its corresponding generated potential / time curve due to a piezoelectric mechanism.

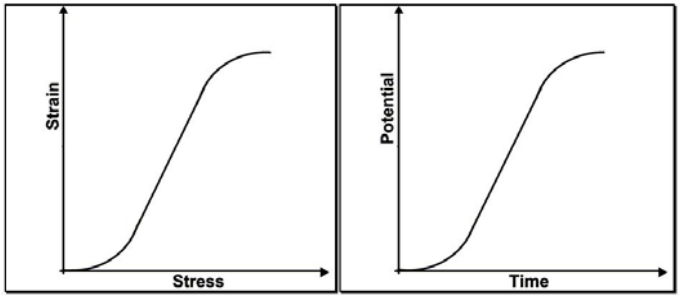

Fig. 22. Typical stress / strain (left) curve and its corresponding generated potential / time (right) curve due to piezoelectricity.

The mechanism of figure (22) is activated in the seismogenic region short before the occurrence of the pending large EQ thus generating the oscillating component of the earth's electric field according to the large scale piezoelectric generating mechanism (Thanassoulas et al. 1986, 1993, 2008) which is presented in figure (23).

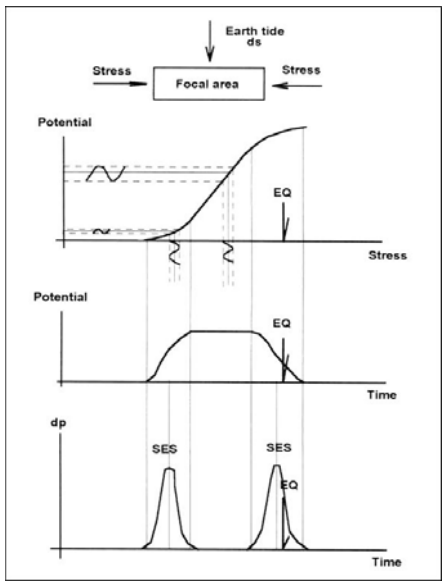

Fig. 23. Large scale piezoelectric mechanism triggered by the tidal waves acting upon the oscillating lithosphere. The case is valid at a seismogenic area some time before the occurrence of the pending large EQ.

This large scale piezoelectric mechanism was validated by what was observed some time before the Skyros EQ being presented in the following figure (24).

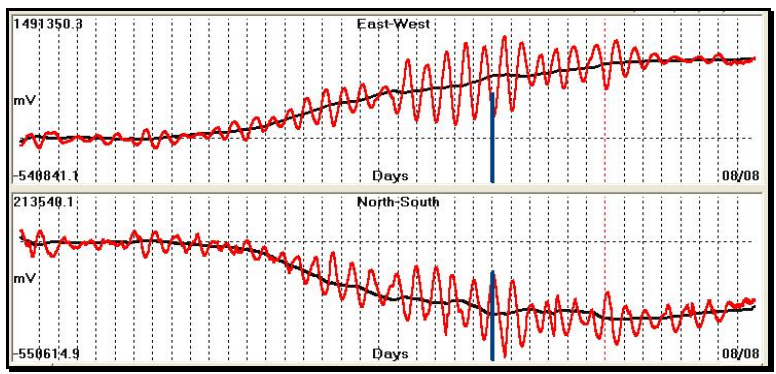

Fig. 24. Oscillating earth's electric field component (red line) on top of the total earth's electric field (black line). Skyros EQ is denoted by a blue bar.

The observed different polarity between the E-W and N-S components is due to the relative location of the monitoring site (VOL) location and the epicenter location of Skyros EQ (Thanassoulas, 2008). It is important to notice that the oscillating electric field increases in amplitude as long as the stress load increases. Close to the end of this phase and when the strain has almost attained its maximum value the large EQ of Skyros takes place while due to stress discharge of the focal area the oscillating component of the earth's electric field decreases in amplitude thereon.

The next figure (25) demonstrates the way the tidal wave of T = 14 days triggers the Skyros EQ.

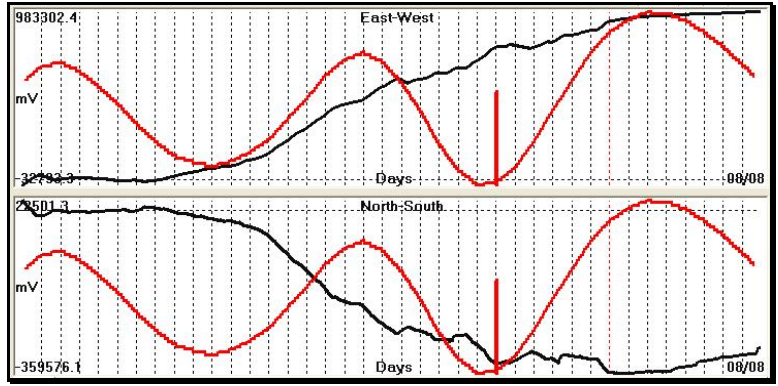

Fig. 25. Correlation of the earth's total field (black line) to the tidal wave of T = 14 days (red line). The Skyros EQ is denoted by a red bar.

During the time period of $25^{th}$ – $26^{th}$ of July 2001, the Skyros seismogenic area had been strained at a level in proximity to the fracturing one. The extra stress load required to trigger the EQ was provided by the tidal wave at the right time. It must be noticed that three earlier tidal peaks did not trigger any EQ just because the seismogenic area had not reached its critical stress level which was attained just when the oscillating component of the earth's electric field started to evolve.

Finally, since the regional seismogenic activated area is known, by applying the Lithospheric Seismic Energy Flow Model on its past years seismic history, it is possible (Thanassoulas, 2001, 2008a,b, Thanassoulas et al. 2001) to determine the stored seismic energy and therefore the maximum expected EQ magnitude (fig. 21). In this case a magnitude of Ms = 6.1R was determined for the Skyros EQ in contrast to the Ms = 5.7R given by the Geodynamic Institute of Athens, Ms = 6.1R given by the Seismic Laboratory of Patras University and Ms = 6.3R given by the seismological observatory of Thessaloniki University. Furthermore, the determined maximum magnitude is indicated by the seismic potential map of Greece calculated for the year 2000 (Thanassoulas et al. 2003, 2008a). A close inspection of the location of the epicenter of Skyros EQ reveals that it is located within the area which is characterized as capable of generating EQs of maximum magnitude of Ms > 6.0R.

Closing this work it is concluded that it is not impossible to shorten the time window for predicting a large EQ to less than an hour. Even better, this time window can be determined some days before the main seismic event. Consequently, there is enough time for the state authorities to take some precaution measures in order to minimize losses of any kind and especially of human beings. Moreover it was demonstrated that the rest of the prognostic parameters (epicenter location and magnitude) is possible to be determined with a quite acceptable for practical purposes accuracy.

In conclusion, the review of the Skyros EQ available data suggests that a really practical "short-term" earthquake prediction procedure is an achievable task with the present day's scientific knowledge. Its future application on other highly seismogenic areas, besides Greece, will prove its validity.

## 4. References

Rudman, J. A., Ziegler, R., and Blakely, R., 1977. Fortran Program for Generation of Earth Tide Gravity Values, Dept. Nat. Res. State of Indiana, Geological Survey.
Thanassoulas, C., 1982. Self potential variations measurements at Skala Kallonis area, Greece, for the period 29/1 – 10/3/1982., Open File report E3648, Inst. Geol. Min. Expl. (IGME), Athens, Greece, pp. 1-16.
Thanassoulas, C., 1991. Determination of the epicentral area of three earthquakes (Ms>6) in Greece, based on electrotelluric currents recorded by the VAN network., Presented in: EUG – VI meeting, Strasbourg, France.
Thanassoulas, C., 1991a. Determination of the epicentral area of three earthquakes (Ms>6) in Greece, based on electrotelluric currents recorded by the VAN network., Acta Geoph. Polonica, Vol. XXXIX, no. 4, pp. 273 – 287.
Thanassoulas, C., 2001. Earthquake prediction based on electrical signals recorded on ground surface., In Seminar Proc. «Possible correlation between electromagnetic earth fields and future earthquakes»., Inst. Nuc. Res. Nuc. Energ., BAS, Sofia, 23-27 July, 2001, Bulgaria.
Thanassoulas, C., 2007. Short-term Earthquake Prediction, H. Dounias & Co, Athens, Greece. ISBN No: 978-960-930268-5
Thanassoulas, C., 2008. "Short-term time prediction" of large EQs by the use of "Large Scale Piezoelectricity" generated by the focal areas loaded with excess stress load. arXiv:0806.0360v1 [physics.geo-ph].
Thanassoulas, C. 2008a. The seismogenic area in the lithosphere considered as an "Open Physical System". Its implications on some seismological aspects. Part – I. Accelerated deformation. http://arxiv.org/0806.4772 [physics.geo-ph]
Thanassoulas, C. 2008b. The seismogenic area in the lithosphere considered as an "Open Physical System". Its implications on some seismological aspects. Part – II. Maximum expected magnitude. http://arxiv.org/0807.0897 [physics.geo-ph]

**Thanassoulas, C., Tselentis, G.A., 1986.** Observed periodic variations of the Earth electric field prior to two earthquakes in N. Greece., In: Proc. 8th European Conf. Earthquake Engineering, Lisbon, Portugal, Vol. 1, pp.41-48.

**Thanassoulas, C., Tselentis, G., 1993.** Periodic variations in the earth's electric field as earthquake precursors: results from recent experiments in Greece., Tectonophysics, 224, 103-111

**Thanassoulas, C., Klentos, V., 2001.** The "energy-flow model" of the earth's lithosphere. Its application on the prediction of the "magnitude" of an imminent large earthquake. The "third paper". IGME, Open file report: A.4384, Greece.

**Thanassoulas, C., Klentos, V. 2003.** Seismic potential map of Greece, calculated by the application of the "Lithospheric energy flow model"., IGME, Open File Report A. 4402, Athens, Greece, pp. 1-25.

**Thanassoulas, C., Klentos, V., Verveniotis, G. 2008.** Multidirectional analysis of the oscillating (T = 24 hours) Earth's electric field recorded on ground surface. arXiv:0808.1801v1 [physics.geo-ph].

**Thanassoulas, C., Klentos, V., 2008a.** The seismogenic area in the lithosphere considered as an "Open Physical System". Its implications on some seismological aspects. Part – III. Seismic Potential. arXiv:0807.1428v1 [physics.geo-ph].